\RequirePackage{fix-cm}
\documentclass[smallextended]{svjour3m}     
\usepackage{nicefrac}
\usepackage{cite}
\usepackage{braket}
\usepackage{graphics}
\usepackage{amsmath}
\usepackage[utf8]{inputenc}
\usepackage{xcolor}
\usepackage[bookmarks=true,bookmarksnumbered=false,bookmarksopen=false, breaklinks=false,pdfborder={0 0 0},backref=false,colorlinks=true,citecolor=red,linkcolor=blue,urlcolor=blue]{hyperref}
\usepackage[b5paper]{geometry}
\newcommand{\tr}{\operatorname{Tr}}
\newcommand{\op}{\boldsymbol}

\newcommand{\orcid}[1]{\href{https://orcid.org/#1}{\resizebox{10px}{!}{\includegraphics{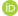}}}}
\journalname{\noindent International Journal of Theoretical Physics
(2022) \\
\href{https://doi.org/10.1007/s10773-022-05030-z}
{\small https://doi.org/10.1007/s10773-022-05030-z} \small{(Published version available on the hyperlink)} \vskip 5mm}
{} \small{}

\begin{document}

\title{Quantifying Entanglement with Coherence}
\author{Neha Pathania\orcid{0000-0002-3385-4761} \and
Tabish Qureshi\orcid{0000-0002-8452-1078}}
\institute{Centre for Theoretical Physics, Jamia Millia Islamia, New Delhi,
India.\\
\email{tabish@ctp-jamia.res.in}}

\maketitle

\begin{abstract}
Quantifying entanglement is a work in progress which is important for the
active field of quantum information and computation. A measure
of bipartite pure state entanglement is proposed here, named
\emph{entanglement coherence},
which is essentially the normalized coherence of the entangled state
in its Schmidt basis. Its value is 1 for maximally entangled states, and
0 for separable states, irrespective of the dimensionality of the
Hilbert space. So a maximally entangled state is also the one which is
maximally coherent in its Schmidt basis. Quantum entanglement and
quantum coherence are thus intimately connected. Entanglement coherence
turns out to be closely related to the \emph{unified entropy} of the reduced
state of one of the subsystems.
Additionally it is shown that the entanglement coherence is closely
connected to the Wigner-Yanase \emph{skew information} of the reduced density
operator of one of the subsystems, in an interesting way.
\end{abstract}


\section{Introduction}
Much after the advent of quantum mechanics, it took a paper by Einstein,
Podolsky and Rosen\cite{epr} to point out its highly nonclassical and
nonlocal nature. The word entanglement came from the observation
by Schr\"odinger that after two independent systems interact, they cannot
generically be described individually by a state of their own\cite{schr}.
Entanglement became an active subject of research because of the
``spookiness" it implied\cite{bell}. Entanglement leads to certain effects
which cannot be simulated by classical systems, a fact which firmly
positions it as an essentially quantum phenomenon\cite{karimi}.
The demonstration of usefulness of
entanglement in quantum cryptography\cite{ekert}, and quantum
teleportation\cite{teleport} led to the development of entanglement as a
subject in itself\cite{rmp}. Huge amount of work has been carried out in
the development of the theory of entanglement. Three broad aspects, which
the research on entanglement deals with, are\cite{rmp} (i) detecting
entanglement theoretically and experimentally, (ii) preventing and possibly
reversing the loss of entanglement in various processes, and
(iii) characterizing and quantifying entanglement. It is this last aspect
which is the subject of the present investigation.

An essential ingredient in the quantification of entanglement is defining
a measure of entanglement. A lot of research effort has already been made
to that end\cite{bruss,plenio2016}, and we can at best only mention some of
the more popular measures in the passing. \emph{Entanglement cost}, denoted
by $E_C(\rho)$ is a measure of how efficiently one can convert a
maximally entangled state to the given state $\rho$\cite{EC1,EC2,EC3}.
\emph{Distillable entanglement}, denoted by $D(\rho)$ is another measure which
deals with the reverse of the process involved in entanglement cost\cite{EC1}.
It deals with how efficiently one can generate a maximally entangled
state from a given state $\rho$. Apart from these two, there are a host
of other entanglement measures based on a different approach to the
problem. To name a few, \emph{entanglement of formation} $E_F(\rho)$ 
\cite{EF}, \emph{relative entropy of entanglement}\cite{REE},
\emph{logarithmic negativity}\cite{N}. We particularly mention a measure
introduced by Hill and Wootters for two qubits, called \emph{concurrence}
\cite{con}, and its generalization to arbitrary Hilbert space dimension,
called \emph{I-concurrence}\cite{rungta,pani}, as we will discuss it in the
context of the result of this work.

An aspect of all the entanglement measures, that we would like to draw 
attention to, is that while they yield value 0 for disentangled states,
there is generally no fixed upper bound for maximally entangled states of
arbitrary Hilbert space dimension. We feel that normalization is a desirable
feature of a good 
entanglement measure because the measure should give an indication of how
close a state is to a maximally entangled state, and also if a given
state is maximally entangled or not. This is, of course, a matter of 
convention, and any entanglement measure can be appropriately normalized.

\section{Coherence of an entangled state}
Let us consider a bipartite system composed of two subsystems $a$ and $b$,
where $a$ has the smaller Hilbert space of the two, in case they are unequal.
Our main result is that for a bipartite entangled state $\ket{\psi}$, its
\emph{entanglement coherence} is the normalized quantum coherence of the
entangled state in its Schmidt basis. It can also be represented in terms
of the reduced density operator of one of two subsystems.
If $\op{\rho}=\tr_b\left(\ket{\psi}\bra{\psi}\right)$
is the reduced density operator for the subsystem $a$, for the bipartite
entangled state $\ket{\psi}$, then its \emph{entanglement coherence} is
given by
\begin{equation}
\mathcal{C}_E = \tfrac{1}{n-1}\left[(\tr\sqrt{\op{\rho}})^2 - 1\right].
\label{ECbi}
\end{equation}
where $n$ is the dimensionality of the Hilbert space of $a$. 

We start by writing a bipartite entangled state $\ket{\psi}$ in its
Schmidt decomposed form\cite{schmidt}
\begin{equation}
\ket{\psi} = \sum_{i=1}^n \sqrt{p_i} \ket{a_i}\ket{b_i},
\label{schmidt}
\end{equation}
where $\ket{a_i}, \ket{b_i}$ represent the basis states of the two
sub-systems, respectively, $n$ is the dimensionality of the smaller
Hilbert space of the two, and $\sqrt{p_i}$ are real and positive numbers
called Schmidt coefficients.
The full density operator for the state (\ref{schmidt}) is given by
\begin{equation}
\rho_F = \ket{\psi}\bra{\psi} = \sum_{j,k=1}^n \sqrt{p_jp_k}
\ket{a_j}\ket{b_j} \bra{a_k}\bra{b_k}.
\label{rhoF}
\end{equation}
Normalized quantum coherence has recently been defined, based on the sum of
absolute values of the offdiagonal elements of the density matrix
\cite{coherence,cvisibility,nduality}. It is a basis-dependent measure.
For the above state, we define the coherence in the Schmidt basis, in
a slightly different way:
\begin{equation}
\mathcal{C}_E \equiv \tfrac{1}{n-1}\sum_{j\neq k} |(\rho_F)_{jk}| = 
 \tfrac{1}{n-1} \sum_{j\neq k} \sqrt{p_jp_k}.
\label{C}
\end{equation}
where $n$ is chosen to be the dimension of the Hilbert space of $a$,
in order to normalize $\mathcal{C}_E$, since we are using a specific
basis here. In the original definition,
coherence is normalized by using $n$ as the dimension of the Hilbert
space of the full system \cite{cvisibility,nduality}. If $a$ and $b$ are
disentangled, only one Schmidt coefficient is nonzero. Consequently,
for disentangled states $\mathcal{C}_E = 0$. The maximally entangled state
is generally defined as the state which has all Schmidt coefficients equal
\cite{plenio2016}.
Thus $\sqrt{p_i}=1/\sqrt{n}, i=1,n$ denotes a maximally entangled state.
It is straightforward to see that for a maximally entangled state,
$\mathcal{C}_E = \tfrac{1}{n-1} \sum_{j\neq k} \frac{1}{n} = 1$. 
The reason for the choice of $n$ in the prefactor $\tfrac{1}{n-1}$ should
be obvious now. A different choice would not give
$\mathcal{C}_E = 1$ for a maximally entangled state.
In general, an entangled state will have $\mathcal{C}_E$
between 0 and 1. This looks like a good measure of entanglement,
and is normalized too.  We can now define
entanglement coherence $\mathcal{C}_E$ as the \emph{normalized coherence of the
entangled state in its Schmidt basis}. It is an elegant measure of entanglement
in that it is always bounded by 0 and 1, irrespective of the Hilbert space
dimension, and is intimately connected to another fundamental property of
quantum states, namely quantum coherence. It is rather satisfying to
observe that maximally entangled states are the ones which are maximally
coherent in their Schmidt basis. It is pertinent to mention that there has
been earlier work which goes in the reverse direction, namely in using
entanglement to measure coherence\cite{beraprl}. Here we use coherence
to measure entanglement. Since entanglement coherence arises from the
Schmidt basis, it is obvious that it is applicable only to pure state
entanglement, and not to mixed entanglement.

The reduced density operator for one of the subsystems is given by
\begin{equation}
\op{\rho} = \tr_b\left(\ket{\psi}\bra{\psi}\right) = \sum_{j=1}^n p_j
\ket{a_j}\bra{a_j} =
\tr_a\left(\ket{\psi}\bra{\psi}\right) = \sum_{j=1}^n p_j
\ket{b_j}\bra{b_j},
\label{rhoa}
\end{equation}
and the density matrix is the same for both.
It would be nice to write $\mathcal{C}_E$ in a basis independent manner.
For that we first notice that since $\op{\rho}$, given by (\ref{rhoa}),
is a diagonal matrix, $\sqrt{\op{\rho}}$ can be simply written as
\begin{equation}
\sqrt{\op{\rho}} = \sum_{j=1}^n \sqrt{p_j} \ket{a_j}\bra{a_j}.
\label{rhosqrt}
\end{equation}
Entanglement coherence $\mathcal{C}_E$, given by (\ref{C}), can now be
manipulated as follows
\begin{eqnarray}
\mathcal{C}_E &=& \tfrac{1}{n-1} \sum_{j\neq k} \sqrt{p_jp_k}
 = \tfrac{1}{n-1} \left(\Big[\sum_{j} \sqrt{p_j}\Big]^2 - \sum_{j} p_j\right) \nonumber\\
 &=& \tfrac{1}{n-1}\left[(\tr{\sqrt{\op{\rho}}})^2 - \tr\op{\rho} \right]
 = \tfrac{1}{n-1}\left[(\tr{\sqrt{\op{\rho}}})^2 - 1 \right].
\end{eqnarray}
Trace being basis independent, one arrives at (\ref{ECbi}).
Thus, for evaluating $\mathcal{C}_E$ one need not go about finding the
Schmidt basis for the entangled state. One can find the reduce density
operator of one subsystem, and employ (\ref{ECbi}) to evaluate $\mathcal{C}_E$.
Notice that the reduced density operator for the subsystem $b$, 
\begin{eqnarray}
\op{\rho'} = \tr_a\left(\ket{\psi}\bra{\psi}\right) = \sum_{j=1}^n p_j
\ket{b_j}\bra{b_j},
\end{eqnarray}
yields the same diagonal matrix in the basis $\{\ket{b_i}\}$, as 
$\op{\rho}$ yields in the basis $\{\ket{a_i}\}$. Then it follows that
\begin{equation}
\mathcal{C}_E = \tfrac{1}{n-1}\left[(\tr\sqrt{\op{\rho}})^2 - 1\right]
= \tfrac{1}{n-1}\left[(\tr\sqrt{\op{\rho'}})^2 - 1\right].
\end{equation}

Notice that $\mathcal{C}_E$ is also a normalized measure of the mixedness
of the reduced density operator $\op{\rho}$. The purity of $\op{\rho}$ implies
$\op{\rho}^2 = \op{\rho}$, which in turn implies
\begin{equation}
\sqrt{\op{\rho}} = \op{\rho} .
\end{equation}
Taking trace of both sides, we arrive at the result
$\mathcal{C}_E \equiv \tfrac{1}{n-1}\left[(\tr\sqrt{\op{\rho}})^2 - 1\right] = 0$.
For a maximally mixed $\op{\rho}$, $\tr\sqrt{\op{\rho}} = \sqrt{n}$, which leads to
$\mathcal{C}_E \equiv \tfrac{1}{n-1}\left[(\tr\sqrt{\op{\rho}})^2 - 1\right] = 1$.

Next we move on to finding the expression for $\mathcal{C}_E$ in an arbitrary
basis of the subsystem $a$.
\begin{eqnarray}
\mathcal{C}_E &=& \tfrac{1}{n-1}\left[(\tr\sqrt{\op{\rho}})^2 - 1\right]
= \tfrac{1}{n-1}\left[(\tr\sqrt{\op{\rho}})^2 - \tr\rho \right] \nonumber\\
&=& \tfrac{1}{n-1} \Big[ \Big(\sum_{i=1}^n\bra{\psi_i}\sqrt{\op{\rho}}\ket{\psi_i}\Big)^2
\nonumber\\
&& - \sum_{i,j=1}^n\bra{\psi_i}\sqrt{\op{\rho}}\ket{\psi_j}
\bra{\psi_j}\sqrt{\op{\rho}}\ket{\psi_i}
\Big] \nonumber\\
&=& \tfrac{1}{n-1} \sum_{i\neq j} \Big[ \bra{\psi_i}\sqrt{\op{\rho}}\ket{\psi_i}
\bra{\psi_j}\sqrt{\op{\rho}}\ket{\psi_j}
- |\bra{\psi_i}\sqrt{\op{\rho}}\ket{\psi_j}|^2\Big]. \nonumber\\
\label{ECb}
\end{eqnarray}
Thus we arrive at the following compact expression for entanglement coherence,
which works for any basis
\begin{equation}
\mathcal{C}_E = \tfrac{1}{n-1} \sum_{j\neq k} \left( \sqrt{\rho}_{jj}\sqrt{\rho}_{kk}
 - |\sqrt{\rho}_{jk}|^2\right).
\label{EC}
\end{equation}
where $\sqrt{\rho}_{jk}$ are the elements of the square-root of the
reduced density matrix in the chosen
basis, and $n$ is the dimensionality of the Hilbert space of $a$.
It is interesting to compare entanglement coherence with the
well known entanglement measure ``I-concurrence." The square of
I-concurrence is given by\cite{pani,rungta}
\begin{equation}
E_a^2 = 2 \sum_{j\neq k} \left( \rho_{jj}\rho_{kk} - |\rho_{jk}|^2\right)
 = 2 \left[1 - \tr(\op{\rho}^2) \right].
\label{IC}
\end{equation}
This expression can be compared with (\ref{EC}) and (\ref{ECbi}).
The factor of 2 in (\ref{EC}) makes sure that the value of $E_a^2$ is
1 for two maximally entangled qubits. However, there is no fixed value
for maximally entangled states in arbitrary Hilbert space dimension.
For the entangled state (\ref{rhoF}), square of I-concurrence is given by
\begin{equation}
E_a^2 = 2 \sum_{j\neq k} p_j p_k
= 2 \sum_{j\neq k} |(\rho_F)_{jk}|^2  .
\label{IC1}
\end{equation}
Now it has been shown that for any density operator $\op{\rho}$, the quantity
$\sum_{j\neq k} |\rho_{jk}|^2$ cannot be a good measure of coherence,
as there exist incoherent operations under which this quantity increases
\cite{coherence}. Here we are not concerned if this quantity is a good measure
of coherence or not, but this result implies that $E_a^2$ given by
(\ref{IC1}) can \emph{increase} under certain \emph{incoherent} operations.
In the light of this result, one may need to reassess if I-concurrence is
a good entanglement measure, and under what kind of incoherent operations
it can increase. 

\begin{figure}
\resizebox{6.5cm}{!}{\includegraphics{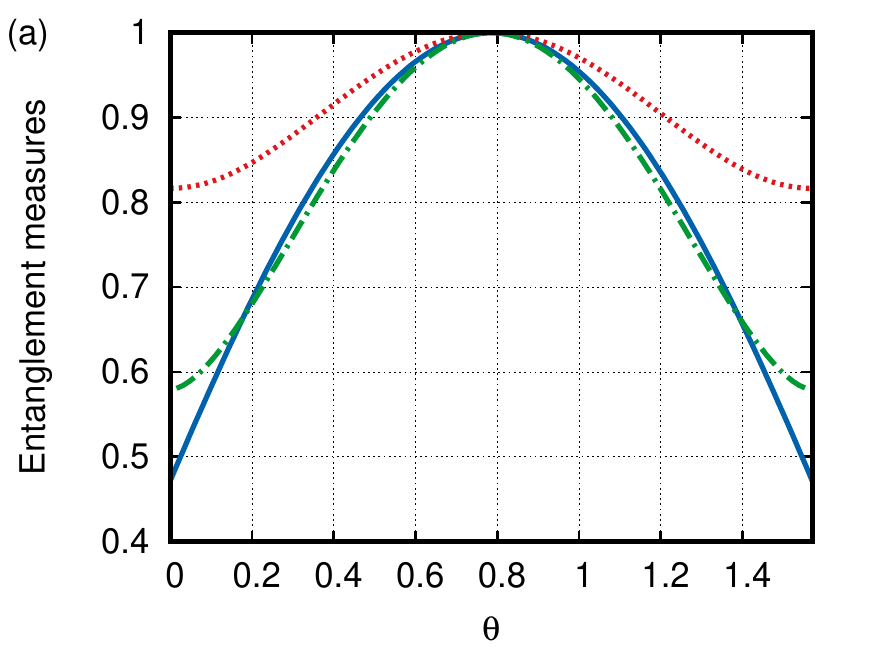}}
\resizebox{6.5cm}{!}{\includegraphics{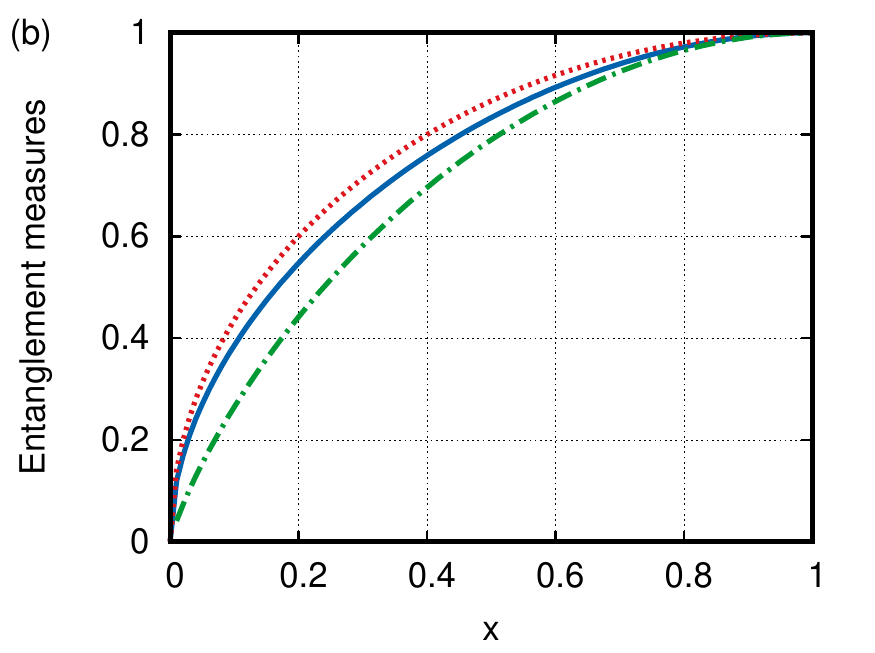}}
\caption{Various entanglement measures plotted for (a) the entangled state
(\ref{state1}) and (b) the entangled state (\ref{state2}):
Entanglement coherence $\mathcal{C}_E$ (solid line), normalized 
I-concurrence (dotted line), and normalized entropy of entanglement 
(dashed line).}
\label{ecplot}
\end{figure}

Although it is difficult to say which entanglement measure is better than the
others, it may be useful to compare the behavior of entanglement coherence
with other measures for some example entangled states. Let us consider the
following bipartite entangled state of two qutrits
\begin{equation} \label{state1}
|\Psi_1\rangle = \sqrt{\tfrac{2}{3}}\sin\theta|+\rangle_1|+\rangle_2
+ \sqrt{\tfrac{2}{3}}\cos\theta|0\rangle_1|0\rangle_2
+ \sqrt{\tfrac{1}{3}}|-\rangle_1|-\rangle_2 ,
\end{equation}
where $|+\rangle,|0\rangle,|-\rangle$ are orthonormal states of a qutrit,
and $0\le\theta\le\pi/2$. The state is maximally entangled for $\theta=\pi/4$.
Another state that can be considered is
\begin{equation} \label{state2}
|\Psi_2\rangle = \sqrt{\tfrac{x}{3}}|+\rangle_1|+\rangle_2
+ \sqrt{\tfrac{x}{3}}|0\rangle_1|0\rangle_2
+ \sqrt{1-\tfrac{2x}{3}}|-\rangle_1|-\rangle_2 ,
\end{equation}
where $0 \le x \le 1$. For $x=0$ the state is disentangled, and for $x=1$
it is maximally entangled.
Entanglement coherence, I-concurrence and entropy of entanglement are plotted
in Fig. \ref{ecplot}.

One may wonder if there is a reason why coherence in the Schmidt basis,
of all the joint basis sets, gives a good measure of entanglement. Here we
try to explore this question. Consider a joint basis for the two subsystems,
$\{|\alpha_i\rangle\otimes\beta_j\rangle\}$. Coherence of the full state
$\rho_F=|\Psi\rangle\langle\Psi|$ in this basis can be written as
\begin{eqnarray}
C_{\alpha\beta}(\rho_F) &=& \sum_{i\ne i'}\sum_{j=j'} |\langle\alpha_i\beta_j|\rho_F|\alpha_{i'}\beta_{j'}\rangle|
+ \sum_{i=i'}\sum_{j\ne j'} |\langle\alpha_i\beta_j|\rho_F|\alpha_{i'}\beta_{j'}\rangle| \nonumber\\
&& + \sum_{i\ne i'}\sum_{j\ne j'} |\langle\alpha_i\beta_j|\rho_F|\alpha_{i'}\beta_{j'}\rangle| \nonumber\\
&=& \sum_{i\ne i'} |\langle\alpha_i|\rho|\alpha_{i'}\rangle|
+ \sum_{j\ne j'} |\langle\beta_j|\rho'|\beta_{j'}\rangle|
 + \sum_{i\ne i'}\sum_{j\ne j'} |\langle\alpha_i\beta_j|\rho_F|\alpha_{i'}\beta_{j'}\rangle ,
\end{eqnarray}
where $\rho$ and $\rho'$ are the reduced density operators of the subsystems
$a$ and $b$, respectively. For a joint basis in which both $\rho$ and $\rho'$
are diagonal, the first two terms in the above expression are zero. Schmidt
basis is precisely that basis. In addition to that, in the third sum
in the above expression, all terms where $i\ne j$ and $i'\ne j'$ are
zero, in the Schmidt basis. So, it may be that, of all the joint basis sets,
the Schmidt basis gives the minimum coherence for the entangled state.

\section{Connection with Unified Entropy}

For a given density operator $\rho$, the \emph{unified entropy},
introduced by Hu and Ye \cite{HuYe},  is defined as 
\begin{equation}
 S^s_r(\rho) = \frac{1}{(1-r)s} \big[(\tr(\rho^r))^s - 1\big], 
\end{equation}
where  $r>0, r\ne 1$, and $s\ne 0$. It is a family of entropies that
yields various quantum entropies as particular or limiting cases. From
(\ref{ECbi}), one can see that if $\rho$ is the reduced density operator
for one of the subsystems of an entangled state, the entanglement coherence
of the entangled state is related to the unified entropy of one of the
subsystems as
\begin{equation}
 \mathcal{C}_E = \frac{1}{n-1} S^2_{\nicefrac{1}{2}}(\rho) = \frac{1}{n-1}
\big[(\tr(\rho^{\nicefrac{1}{2}}))^2 - 1\big]. 
\end{equation}
The reduced state of one of the two subsystems is expected to have a
non-zero von Neumann entropy, if the state of the system is entangled.
However, it is interesting to observe that the coherence of the entangled
state in its Schmidt basis is virtually the same as a particular case of the
unified entropy of the reduced state of one of the subsystems. Hu and Ye have
presented several results for various properties of unified entropy
\cite{HuYe}, many of which will apply to entanglement coherence too.

\section{Connection with skew information}

In 1963 Wigner and Yanase introduced the concept of \emph{skew information},
which is believed to quantify the \emph{quantum} part of the uncertainty of
an observable in a \emph{mixed state}\cite{WY}. For a mixed state, the usual
measure of uncertainty, the variance, incorporates both quantum and classical
uncertainty.  The skew information of an observable
$A$, in a state $\rho$ is defined as\cite{WY}
\begin{equation}
I(\rho,A) := -\tfrac{1}{2}\tr{[\sqrt{\rho},A]^2}
= \tr{\rho A^2} - \tr{\sqrt{\rho}A\sqrt{\rho}A},
\label{skew}
\end{equation}
where the square brackets denote the commutator. The skew information is
useful when $\rho$ is mixed. For a pure state ($\sqrt{\rho}=\rho$) the
skew information reduces to the usual \emph{variance}
$\langle A^2\rangle - \langle A\rangle^2$. Skew information has emerged
as an interesting and useful tool\cite{luo,banik}.

\subsection{Basis optimization}

For mixed states, one may want to know what is the maximum coherence
of the state, for any basis. To investigate the basis dependence,
one can define a coherence based on the Wigner-Yanase skew information,
specific to a basis. Then one can ask which basis maximizes it \cite{CI}.
Given a basis
$\{|k\rangle\}$ for the subsystem $a$, we can calculate the skew information
for a projection operator $|k\rangle\langle k|$ as
\begin{equation}
I(\rho,|k\rangle\langle k|) = -\tfrac{1}{2}\tr{[\sqrt{\rho},|k\rangle\langle k|]^2},
\label{skewk}
\end{equation}
where $\rho$ is the reduced density operator of (say) the subsystem $a$.
We define a \emph{skew-information based coherence} as
\begin{equation}
\mathcal{C}_I = \sum_{k=1}^{n}I(\rho,|k\rangle\langle k|) ,
\label{CI}
\end{equation}
where the sum goes over all states of the basis. Because of the sum, this
quantity depends on the whole basis. The skew information normally depends
on one particular observable. Here we can consider the reduced density 
operator of our entangled state $\rho$. So, this quantity can be treated as a
basis dependent measure of coherence of the state $\rho$. Next we ask,
which basis maximizes this
coherence, i.e., maximizes the sum of skew informations.
This skew-information based coherence can be written as \cite{CI}
\begin{eqnarray}
\max_{\{|k\rangle\}}\mathcal{C}_I &=&  \max_{\{|k\rangle\}}
\sum_{k=1}^{n}\Big(\tr{(\rho(|k\rangle\langle k|)^2)} - \tr{(\sqrt{\rho}|k\rangle\langle k|\sqrt{\rho}|k\rangle\langle k|)}\Big) \nonumber\\
&=& \max_{\{|k\rangle\}}\sum_{k=1}^{n}(\langle k|\rho|k\rangle - \langle k|\sqrt{\rho}|k\rangle^2) \nonumber\\
&=& 1 - \min_{\{|k\rangle\}}\sum_{k=1}^{n} \langle k|\sqrt{\rho}|k\rangle^2 \nonumber\\
&=&  1 - \min_{\{|k\rangle\}}\sum_{k=1}^{n} \langle k|\sqrt{\rho}|k\rangle^2 \nonumber\\
&=& 1 - \tfrac{1}{n}\Big(\sum_{i=1}^{n} \sqrt{p_i}\Big)^2  ,
\end{eqnarray}
where $p_i$ are the eigenvalues of $\rho$. Since $\sum_{i=1}^{n} \sqrt{p_i}
= \tr{\sqrt{\rho}}$, the above result can written as
\begin{eqnarray}
\max_{\{|k\rangle\}}\mathcal{C}_I &=& 1 - \tfrac{1}{n}(\tr{\sqrt{\rho}})^2
 = \tfrac{n-1}{n}(1 - \mathcal{C}_E) \nonumber\\
\mathcal{C}_E &=& 1 - \tfrac{n}{n-1} \max_{\{|k\rangle\}}\mathcal{C}_I .
\end{eqnarray}
This result says that the entanglement coherence of an entangled state is
closely related to optimal skew-information based coherence of the reduced
state of the subsystem $a$. Intuitively, stronger the entanglement between
the two subsystems, smaller will be any measure of coherence of the reduced
state of one of the subsystems. The skew-information based coherence, of one
of the subsystems, has a closer connection with the entanglement of the
composit system.

\subsection{Quantum uncertainty}

In the following we connect the entanglement coherence to another
quantity based on the skew information, which has been studied earlier.
Now, the skew information given by (\ref{skew}) is dependent on the
observable $A$. If one is looking for an inherent property of the state,
which is not tied down to one observable, one can carry out the following
procedure. If the Hilbert space of the system is n-dimensional, the
set of all observables form another Hilbert space which is
$n^2-$dimensional, if one defines the inner product of two observables
$A$ and $B$ as $\langle A,B\rangle := \tr{AB}$. One can define a basis
of observables on this $n^2-$dimensional Hilbert space as
$X_1, X_2, X_3,\dots , X_{n^2}$. A \emph{quantum} uncertainty, for a mixed
state $\rho$, can then be defined as the sum of the skew informations of
all these $n^2$ observables\cite{luo2}
\begin{equation}
Q(\rho) := \sum_{i=1}^{n^2} I(\rho,X_i).
\label{Qdef}
\end{equation}
It can be shown that $Q(\rho)$ is independent of the particular basis
$\{X_i\}$. Consequently this quantum uncertainty for a mixed
state can be evaluated to yield\cite{luo2}
\begin{equation}
Q(\rho) = \sum_{i=1}^{n^2} I(\rho,X_i) = n - (\tr{\sqrt{\rho}})^2 .
\label{Q}
\end{equation}
Using (\ref{ECbi}) and (\ref{Q}), one then finds
\begin{equation}
\mathcal{C}_E = 1 - \tfrac{1}{n-1}Q(\rho) .
\label{ECQ}
\end{equation}
Thus one finds that the entanglement coherence of an entangled state is
closely connected to the skew information based \emph{quantum} uncertainty
of the reduced density operator of one of the subsystems.

In general, the uncertainty of a mixed state has two quite different origins.
One is classical mixing, and the other is quantum randomness. The
\emph{quantum} uncertainty $Q(\rho)$ is supposed to represent the latter.
If two subsystems are maximally entangled, the reduced state of one of the
two is not expected to have any quantum part of uncertainty. The
subsystem will look like a classically mixed state. On the other extreme,
if the two subsystems are disentangled, the reduced state of one of the
two will be pure, and will have maximal quantum uncertainty, and
no classical mixedness. The relation (\ref{ECQ}) quantifies this 
connection between entanglement and the quantum uncertainty of one
of the subsystems.

\section{Discussion and Conclusion}

One might ask if the measure, entanglement coherence, which is defined
for pure entangled states, can be extended to mixed entangled states.
One way to extend the definition is by a convex roof construction \cite{rmp}
\begin{equation}
\mathcal{C}_{m}(\varrho) = \min_{\{q_k,|\Psi_k\rangle\}}
\sum_k q_k \mathcal{C}_E(|\Psi_k\rangle) ,
\end{equation}
where $\{q_k,|\Psi_k\rangle\}$ is a decomposition of the mixed state density
operator $\varrho$ to pure states
$\varrho = \sum_k q_k |\Psi_k\rangle\langle\Psi_k|$, and
$\mathcal{C}_E(|\Psi_k\rangle)$ is the entanglement coherence corresponding
to the pure entangled state $|\Psi_k\rangle$. A method of evaluating such
convex roof entanglement measures was recently proposed \cite{guhne}.

In conclusion, we have shown that normalized coherence of an entangled
bipartite state, in the Schmidt basis, can be considered a good measure
of the degree of entanglement of the two systems. It uncovers an interesting
connection between and entanglement and coherence, which are separately
studied interesting properties of quantum systems. The normalized coherence
of an entangled state is also interestingly connected to the \emph{quantum}
uncertainty and optimal coherence of the reduced density operator of one
of the subsystems, defined through Wigner-Yanase skew information. It is
also intimiately connected to the unified entropy.
Ramifications of these connections may be explored further.

\section*{Acknowledgments}
Neha Pathania acknowledges financial support from the Department of Science
and Technology, through the Inspire Fellowship (registration code IF180414).







\end{document}